\documentclass[fleqn,usenatbib]{mnras}

\usepackage{newtxtext}

\usepackage[T1]{fontenc}

\DeclareRobustCommand{\VAN}[3]{#2}
\let\VANthebibliography\thebibliography
\def\thebibliography{\DeclareRobustCommand{\VAN}[3]{##3}\VANthebibliography}

\usepackage{graphicx}	
\usepackage{amsmath}	
\usepackage{romannum}
\usepackage{caption}
\usepackage{float}
\restylefloat{table}
\newrobustcmd{\B}{\bfseries}

\newcommand{\xb}{XB~1254$-$690}
\newcommand{\astrosat}{\textit{AstroSat}}

\title[\astrosat{} Observations of \xb]{\astrosat{} Observations of the Dipping Low Mass X-ray Binary \xb{}}

\author[Navale et al.]{
Nilam R. Navale$^{1,2}$\thanks{E-mail: nil09navale@gmail.com},
Devraj Pawar$^{2}$,
A. R. Rao$^{3,4}$,
Ranjeev Misra$^{4}$,
Sudip Chakraborty$^{5}$, 
Sudip Bhattacharyya$^{3}$
\newauthor
and Vaishali A. Bambole$^{1}$
\\
$^{1}$University Department of Physics, University of Mumbai, Kalina, Santacruz, Mumbai 400 098, India\\
$^{2}$Ramniranjan Jhunjhunwala College, Ghatkopar, Mumbai 400086, India\\
$^{3}$Department of Astronomy and Astrophysics, Tata Institute of Fundamental Research, Mumbai 400005, India\\
$^{4}$Inter-University Center for Astronomy and Astrophysics, Ganeshkhind, Pune 411007, India\\
$^{5}$Universit\'e Paris Saclay, Universit\'e Paris Cit\'e, CEA, CNRS, AIM, F-91191 Gif-sur-Yvette, France
}

\date{Accepted 2024 July 3. Received 2024 July 3; in original form 2022 August 16}

\pubyear{2022}

\begin{document}
\label{firstpage}
\pagerange{\pageref{firstpage}--\pageref{lastpage}}
\maketitle

\begin{abstract}
\xb{} is a neutron star low-mass X-ray binary with an orbital period of 3.88 hrs, and it exhibits energy-dependent intensity dips, thermonuclear bursts, and flares. We present the results of an analysis of a long observation of this source using the \astrosat{} satellite. The X-ray light curve gradually changed from a high-intensity flaring state to a low-intensity one with a few dips. The hardness intensity diagram showed that the source is in a high-intensity banana state with a gradually changing flux. Based on this, we divide the observation into four flux levels for a flux-resolved spectral study. The X-ray spectra can be explained by a model consisting of absorption, thermal emission from the disc and non-thermal emission from the corona. From our studies, we detect a correlation between the temperature of the thermal component and the flux and we examine the implications of our results for the accretion disc geometry of this source. 
\end{abstract}




\section{Introduction} \label{sec:intro}

Low Mass X-ray Binary sources containing neutron stars (NS-LMXBs) are known for their variable nature. In X-rays, they exhibit dips, Quasi-Periodic Oscillations (QPOs), bursts, flares and coherent millisecond-period X-ray pulsations. Dips in soft X-rays are seen in a few NS-LMXBs (called the dippers), and in a few black hole LMXBs \citep{2014A&A...564A..62D}. These dips are thought to be caused by the periodic obscuration of X-rays by the accretion stream when it strikes the structures above the accretion disc. Such obscurations are possible only if the LMXBs are high inclination systems so that the line of sight passes through these structures \citep{1982ApJ...253L..61W}. The dips may recur at the system's orbital period, and the change in the X-ray spectrum during a dip is complex. The structure and location of emitting and absorbing components in NS-LMXBs can be well studied by modelling these spectral changes. The dippers sometimes exhibit low frequency ($\sim$1 Hz) quasi periodic oscillations (QPOs), called ``dipper QPOs'', at low intensities \citep{1999ApJ...516L..91H,1999ApJ...511L..41J,2000ApJ...531..453J,2006ApJ...644.1085B}. Hence, dippers allow us to constrain the properties of the accretion discs and the polarized plasma above them \citep{2003ApJ...590..432J} and the QPOs help us understand the innermost region of the accretion disc  \citep{2000ApJ...531..453J}.
 
Another X-ray variability that is mainly observed in NS-LMXBs is X-ray bursts. X-ray bursts are characterized by an increase in the intensity of the emission of X-rays with a short rise time of about a few seconds and a longer decay time. Unstable nuclear burning on the neutron star's surface could lead to such X-ray bursts \citep{2010csxs.book.....L}. In the past few years, burst oscillations have been observed from several X-ray sources. The discovery of burst oscillations at the spin frequencies of the accreting millisecond pulsars SAX J1808.4–3658 and XTE J1814–338 \citep{2003Natur.424...42C,2003ApJ...596L..67S} has conclusively linked these oscillations to neutron star spin. Indeed, the detection of burst oscillations has facilitated the spin frequency measurements of neutron stars in more than a dozen NS-LMXBs. Detection of a low spin frequency neutron star from NS-LMXBs can be beneficial. Besides, in addition to measuring spin rates, the modelling of burst oscillation light curves can be necessary for the measurements of stellar mass and radius \citep{2005ApJ...619..483B}, and for understanding the thermonuclear flame spreading on neutron stars \citep{2006ApJ...642L.161B}.

X-ray flaring in NS-LMXBs has been known for many years. It was first observed in Sco X-1, which took place during both quiescent and active states of the source \citep{1975ApJ...197..457C}. \cite{1985ApJ...296..475W} observed a notable increase in the black body temperature during flaring along with an increase in the emitting area. Later it was observed that many other bright NS-LMXBs exhibit flares similar to Sco X-1. Nevertheless, not much work has been done to develop a physical understanding of the flaring phenomenon. The alteration between flaring in brighter sources and bursting in less bright sources is also not understood. \cite{2001A&A...378..847B} reported the study of flaring in the big dipper X 1624-490. They proposed that the flaring consists of the superposition of X-ray bursts which occurs rapidly because of the higher ($\sim$10 times) mass accretion rate. 

\xb{} is an NS-LMXB showing several of the above features. It is a dipping source and has a 19$^{th}$ magnitude blue star called GR Mus as an optical counterpart \citep{1978Natur.276..247G}. The detection of Type \Romannum{1} X-ray bursts confirmed that the compact object in this source is a neutron star \citep{1980Natur.287..516M}. \xb{} is known for exhibiting energy dependant X-ray dips with an orbital period of 3.88 $\pm$ 0.15 hrs \citep{1986ApJ...309..265C}. \cite{1987ApJ...313..792M} reported the modulation in the optical light curve with the period same as the X-ray dips, with a temporal offset of 0.2 cycles after the X-ray dips. The presence of dips and the absence of eclipse in this source lead to constraining the inclination angle, \textit{i}, to be between 65$^{\circ}$ and 73$^{\circ}$ \citep{1986ApJ...309..265C,1987ApJ...313..792M}.

While studying the timing properties of \xb{}, \cite{2007MNRAS.377..198B} established that it is an atoll source by calculating the color-color diagram using \textit{RXTE} observation and correlating the timing properties with the different sections of it. The authors also mentioned that the source was always found in the banana state and never found in low intensity spectrally hard island state. The first evidence of QPOs from \xb{} was presented by \cite{2011MNRAS.411.2717M} at frequencies 48.63 Hz and 64.01 Hz. The suggestive evidence of 95 Hz oscillations from a thermonuclear burst from \xb{} indicated that the central compact object is spinning with a frequency of 95 Hz \citep{2007MNRAS.377..198B}. However, further confirmation of both QPOs and thermonuclear burst oscillations is yet to be done.  

\xb{} has shown a typical variation in the dip depth in the earlier observations, which makes it different from other dipping NS-LMXBs. \textit{Ginga} observations of the source in 1990 showed deep dips in the light curve \citep{1997PASJ...49..353U}. However, dips were absent during the \textit{BeppoSAX} observations in 1998 \citep{2001ApJ...548..883I} and in \textit{RXTE} observations in 1997. In 1997 optical observations revealed that the amplitude of the optical variability was reduced by $\Delta$\textit{V} $\sim$0.1 mag with mean \textit{V} magnitude unchanged \citep{1999ApJ...527..341S}. Non-detection of dips indicated that the angular size of the disc edge decreased to less than 10$^{\circ}$ from between 17$^{\circ}$ to 25$^{\circ}$ \citep{1999ApJ...527..341S}. The dips were detected again during the \textit{XMM-Newton} observation in January 2001 and during \textit{RXTE} observation in May 2001 and were again undetected in the December 2001 observation \citep{2002ApJ...581.1286S}, in February 2002 observation \citep{2003A&A...407.1079B} and in October 2003 observation \citep{2007A&A...464..291I}. Once again, the deep dipping was observed in the May 2004 \textit{RXTE} observation of the source \citep{2007MNRAS.380.1182B}.

Along with energy-dependent intensity dips and thermonuclear X-ray bursts, the source also exhibits flares. \cite{2002ApJ...581.1286S} reported flaring along with dips (see figure 1 of \cite{2002ApJ...581.1286S}) and bursts (see figure 2 of \cite{2002ApJ...581.1286S}).

In this paper, we present the results of an analysis of a long observation of the dipping NS-LMXB \xb{} using \astrosat{} carried out between 2018 May 03 and 2018 May 05. We observed flares and dips in the X-ray light curve during this observation. However, we did not detect any thermonuclear bursts. We studied the X-ray spectra of this dipping source using data from two instruments on board \astrosat, namely Soft X-ray Telescope (SXT) and Large Area X-ray Proportional Counter (LAXPC), using detailed spectral models. The structure of the paper is as follows. In section \ref{sec:obs_analysis} we describe the data reduction procedure of both SXT and LAXPC instruments. In section \ref{subsec:lxp_hid}, we describe the HID generation and formation of four sections for flux resolved spectroscopy based on intensity. We present a detailed spectral analysis of \astrosat{} data in section \ref{subsec:spec_analysis} along with details of constraining disc inner edge radius in section \ref{subsec:inneredgerad}. Finally, we epitomize the results and discuss their implications in section \ref{sec:discussion}.

\section{Observation and Analysis} \label{sec:obs_analysis}

The Indian astronomical observatory \astrosat{} \citep{2006AdSpR..38.2989A,2014SPIE.9144E..1SS} observed \xb{} between 2018 May 03 and 2018 May 05 (ObsID: 9000002074). We analysed the data from two instruments onboard \astrosat{} namely SXT and LAXPC. The details of the observation with exposure time for both SXT and LAXPC are given in the table \ref{Tab:1}. 

\begin{table}[ht]
   \centering
   \caption{\astrosat{} Observation Details} 
    \label{Tab:1}
    \begin{tabular}{ccc}
    \hline\noalign{\smallskip}
    Obs ID     & Obs. time (UT)    & Exposure (s) \\
           & (yyyy-mm-dd)      &   \\
    \noalign{\smallskip}\hline\noalign{\smallskip}
    9000002074 & 2018-05-03 &
    \begin{tabular}{c} 40700 (SXT) \\99750 (LAXPC)
    \end{tabular} \\
    \noalign{\smallskip}\hline
   \end{tabular}
\end{table}

\subsection{SXT} \label{subsec:sxt_analysis}
The Soft X-ray Telescope \citep[SXT;][]{2016SPIE.9905E..1ES,2017JApA...38...29S} on board \astrosat{} was the primary instrument for the observation, and the data were acquired in the Photon Counting (PC) mode. The source light curve and spectrum are extracted using the \textit{XSELECT} tool in \textit{HEASoft} (version 6.26.1) from $14'$ circular region, which includes $\sim$94$\%$ of the total source photons and is suitable for the lowest possible systematic. For the background light curve, we consider a $14'-16'$ annular region. For spectral analysis, we use the response matrix file (RMF), deep blank sky background spectrum and an auxiliary response file (ARF) provided by the SXT payload operation team\footnote{\url{https://www.tifr.res.in/~astrosat_sxt/dataanalysis.html}}.

\subsection{LAXPC} \label{subsec:lxp_analysis}
The Large Area X-ray Proportional Counter \citep[LAXPC,][]{2016ApJ...833...27Y} data are analysed using the tools from \textit{LAXPCSoft}\footnote{\url{http://astrosat-ssc.iucaa.in/uploads/laxpc/LAXPCsoftware_Aug4.zip}}. By the time of the present observation, LAXPC10 showed unforeseeable variations in HV, and LAXPC30 suffered a gas leakage which caused gain instability. Hence, we are using only LAXPC20 data for spectral studies. However, we use both LAXPC10 and LAXPC20 for timing studies as the HV variations do not significantly affect the power density spectra (PDS). The background is dominating beyond 20.0 keV in LAXPC hence, we consider the 4.0 -- 20.0 keV energy range and extract photons from the top layer of the unit to achieve a better signal-to-noise ratio (S/N).

\subsection{Light curve} \label{subsec:sxt_lxp_lc}
\begin{figure*}
    \centering
    \includegraphics[scale=0.5]{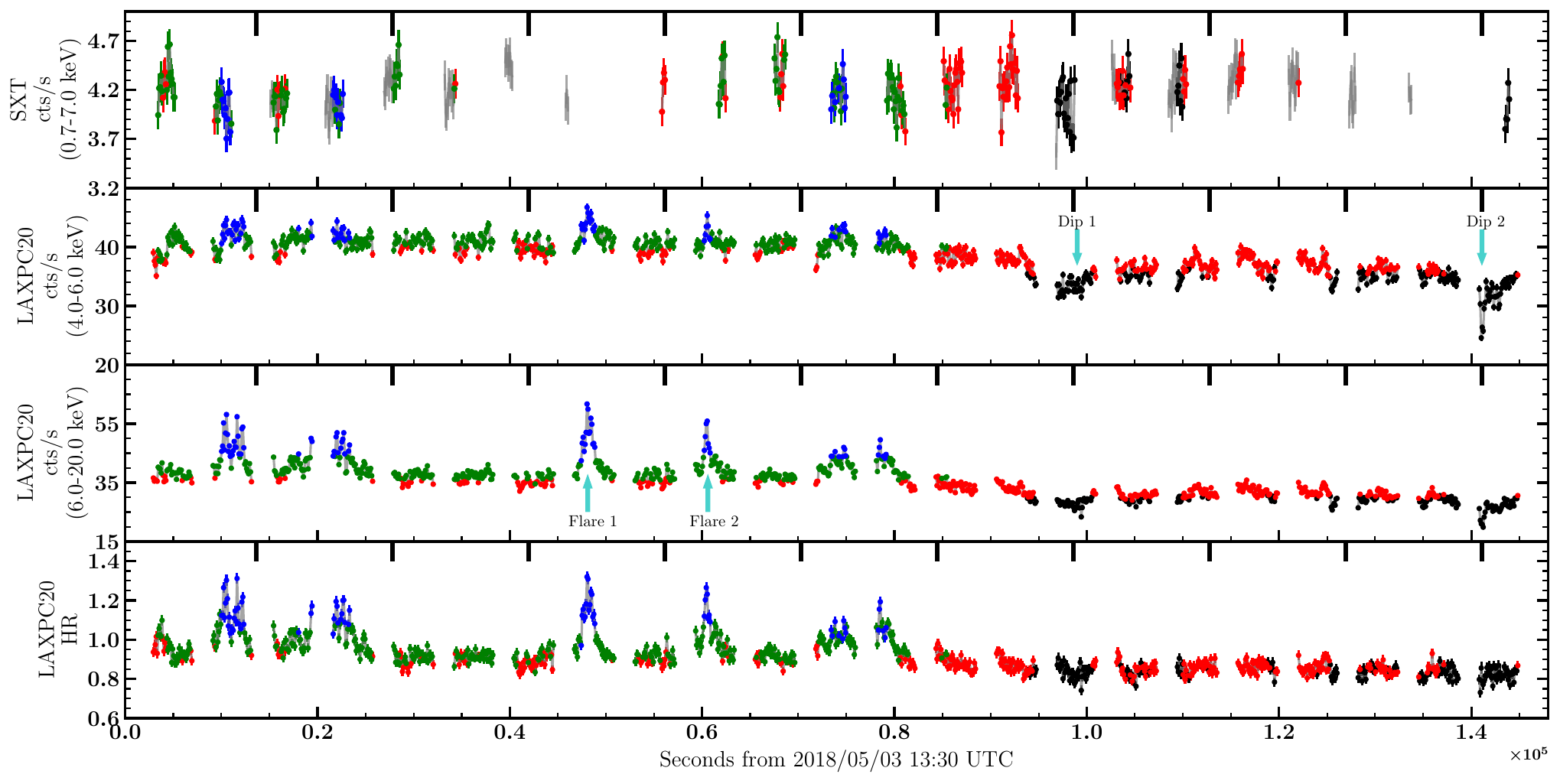}
    \caption{0.7 -- 7.0 keV \astrosat{}/SXT background-corrected light curve (first panel) followed by \astrosat{}/LAXPC20 background-corrected light curves from the top layer of the unit in two different energy bands (second panel: 4.0 -- 6.0 keV, third panel: 6.0 -- 20.0 keV) for \xb{}. The fourth panel is the hardness ratio using two energy bands of LAXPC20. All the light curves and hardness ratio are binned over 100 s. The colors black, red, green and blue in the figure represent the four intensity-dependent sections we are considering for the flux resolved spectroscopy in this work (see section \ref{subsec:lxp_hid} for details). The grey data points represent the total observation. The expected dip centres derived from the folding period and the corrected reference epoch are indicated with black thick vertical tick marks on the upper X-axis of all four panels (see section \ref{subsec:sxt_lxp_lc} for details). The observed two dips are pointed with the arrows in the second panel, and two prominent flares are pointed with the arrows in the third panel.}
    \label{fig:fig1}
\end{figure*}

Figure \ref{fig:fig1} shows the background-subtracted light curves with a bin size of 100 s in different energy ranges: 0.7 -- 7.0 keV (SXT, top  panel), 4.0 -- 6.0 keV (LAXPC20, second panel), 6.0 -- 20.0 keV (LAXPC20, third panel). The fourth panel in the figure shows the hardness ratio (counts in the 6.0 -- 20.0 keV energy range divided by those between the 4.0 -- 6.0 keV band). The count rate in the SXT light curve is normalised with respect to the detection area. The data points in the figure are coded in four different colors (black, red, green and blue), representing the sections we are considering for flux-resolved spectroscopy. The grey data points represent the total observation. These intensity-dependent sections are discussed in section \ref{subsec:lxp_hid} in detail. 
\citet{2017RAA....17..108G} reported the new linear orbital ephemeris (see equation 5 in \citep{2017RAA....17..108G}) with 11931.8069(16) TJD as a newly corrected reference epoch and 14160.004(6) s orbital period. The thick vertical tick marks on the top X-axis of each panel represent the expected dip times using this ephemeris.



\xb{} shows a decrease in its intensity during the present observation. Interestingly, several flare-like features are seen at the beginning of the observation, particularly in the first half. These flares are prominently seen in the 6.0 -- 20.0 keV energy band of LAXPC20 light curve (pointed with the arrows in the third panel of Figure \ref{fig:fig1}) and in the hardness ratio plot (fourth panel of Figure \ref{fig:fig1}). We detected two dips after the flaring episode towards the end of the observation. Both are evident in the 4.0 -- 6.0 keV energy band of LAXPC20 (pointed with the arrows in the second panel of Figure \ref{fig:fig1}). To characterise dips we followed the definition of ``deep'', ``shallow'' and ``undetected'' dips given in \cite{2009A&A...493..145D}. The \astrosat{} data have gaps due to Earth occultation and South Atlantic Anomaly (SAA). While calculating the reduction of the flux during dips compared to the persistent emission, we considered the light curves between these data gaps. The dip 1 is a ``shallow'' dip (near 1.0 $\times$ $10^{5}$ s) with 9.7\% reduction of 4.0 -- 20.0 keV flux and the dip 2 is a ``deep'' dip (near 1.4 $\times$ $10^{5}$ s) with 15.9\% reduction in the flux when compared to the persistent emission. The appearance of these dips coincides with the expected dip times derived from the new linear orbital ephemeris given by \citet{2017RAA....17..108G}

the reference epoch of deep dip observed during \textit{XMM-Newton} observation \citep{2009A&A...493..145D}. 

\subsection{Hardness Intensity Diagram} \label{subsec:lxp_hid}
\begin{figure}
    \centering
  \includegraphics[width=1.0\linewidth]{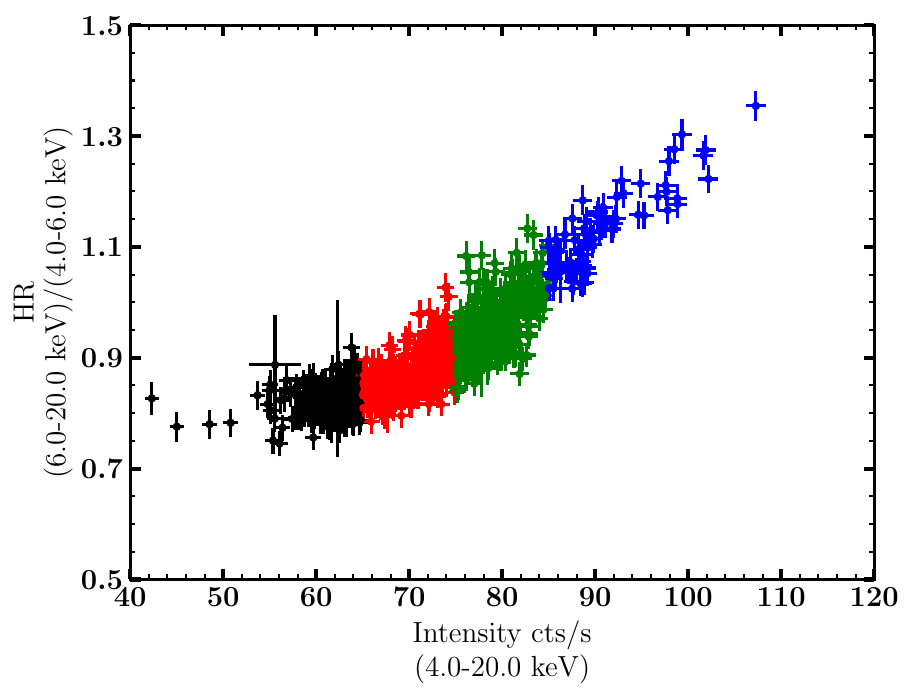}
  \caption{Hardness Intensity Diagram (HID) of \xb{} using \astrosat{}/LAXPC20 data from top layer with 100 s binning. The source was in the banana branch during the observation. HID is divided into four sections based on intensity to perform flux resolved spectroscopy. The flares belong to the blue section of HID, where both intensity and hardness ratio are high, and dips belong to the black section of HID, where intensity and hardness both are low.}
    \label{fig:fig2}
\end{figure}

The Hardness Intensity Diagram (HID) for \xb{} is generated using LAXPC20 data from the top layer of the detector (Figure \ref{fig:fig2}). The hardness ratio is the ratio of background-corrected counts between 6.0 -- 20.0 keV to the counts between 4.0 -- 6.0 keV, and the intensity is the total background-corrected counts between 4.0 -- 20.0 keV. 

Based on the intensity, we divided the HID into four sections (marked in different colors in Figure \ref{fig:fig2}) to perform flux resolved spectroscopy of \xb{}. The selection of sections is made so that each section will have at least 10000 data points. 
We used \textit{laxpc\_fluxresl\_gti} tool from \textit{LAXPCSoft} to generated Good Time Intervals (GTIs) for these sections. The GTIs generated from this tool are then used to generate both SXT and LAXPC20 spectra.

Figure \ref{fig:fig2} distinctly indicates that both intensity and hardness are high during flaring (see the blue points in Figure \ref{fig:fig1}) as flaring is high energy phenomenon and are low during the dips (see the black points in Figure \ref{fig:fig1}) as dipping is a low energy phenomenon.  

\subsection{Dips and Flares Light curve}
\begin{figure}
    \centering
    \includegraphics[width=1.0\linewidth]{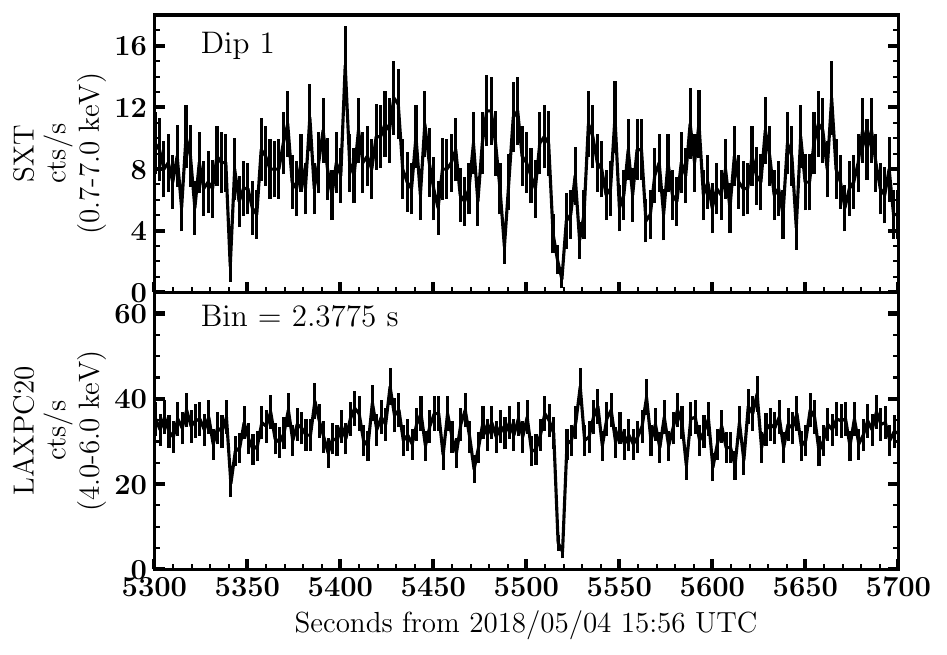}
    \caption{\astrosat{}/SXT 0.7 -- 7.0 keV light curve (upper panel) and \astrosat{}/LAXPC20 4.0 -- 6.0 keV light curve (lower panel) for dip1 (shallow dip) observed in \xb{} with time bin of 2.3775 s.}
    \label{fig:fig3}
\end{figure}

The shallow dip is observed in both SXT and LAXPC20, while the deep dip is observed only in the LAXPC20 data as SXT have a data gap during this time. Figure \ref{fig:fig3} shows the SXT and LAXPC20 background-corrected light curve for the shallow dip with the time bin 2.3775 s (a minimum time resolution of SXT). 

\begin{figure*}
    \centering
    \includegraphics[width=0.45\linewidth]{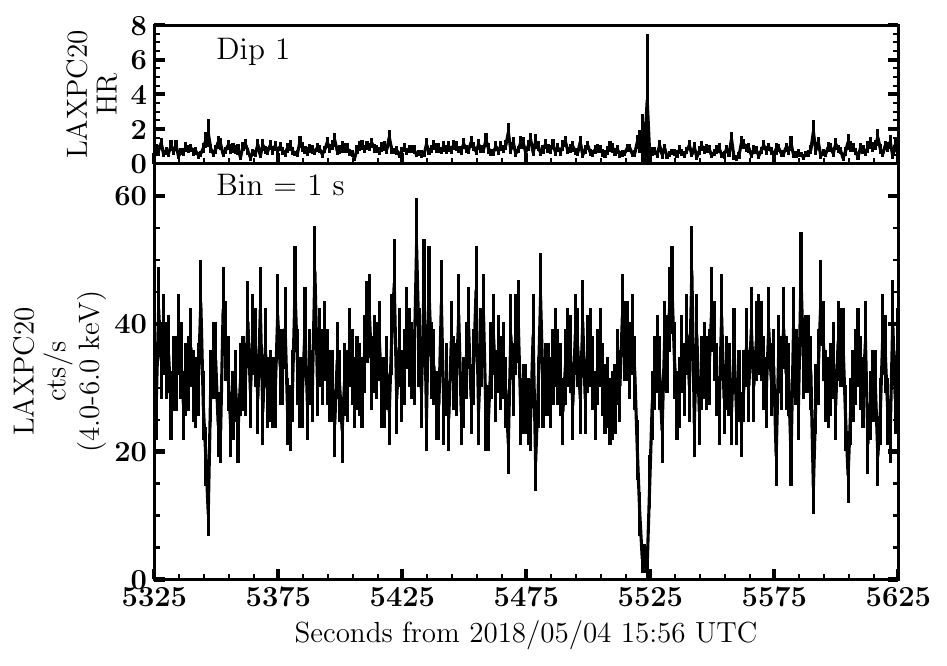}
    \includegraphics[width=0.45\linewidth]{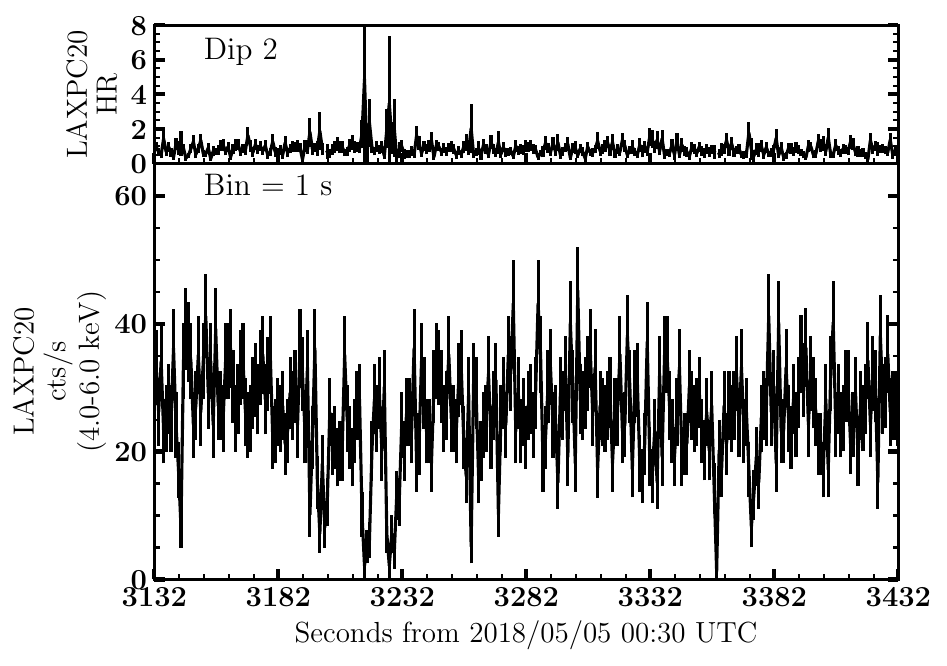}
    \includegraphics[width=0.45\linewidth]{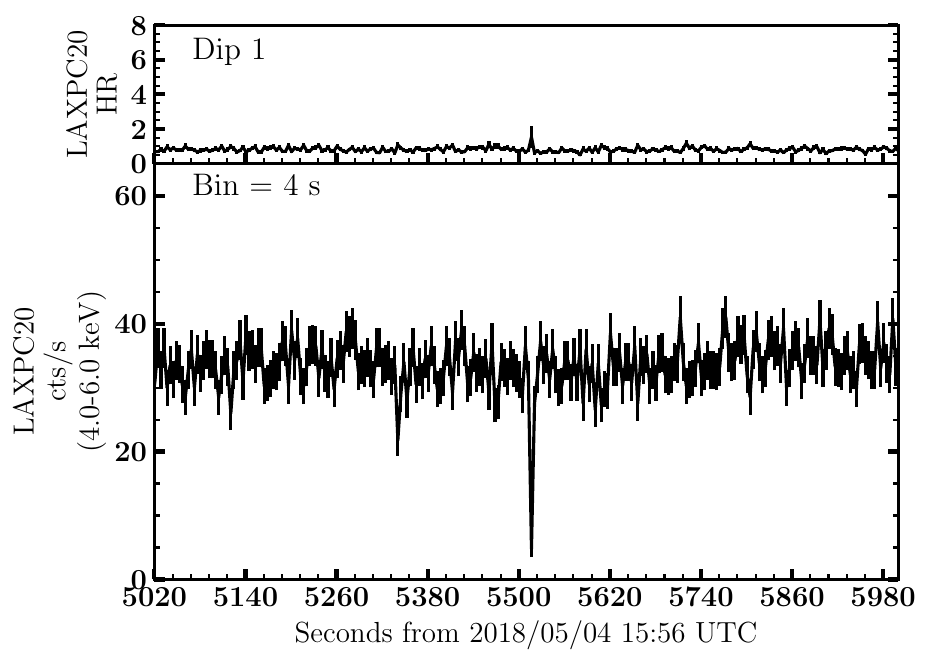}
    \includegraphics[width=0.45\linewidth]{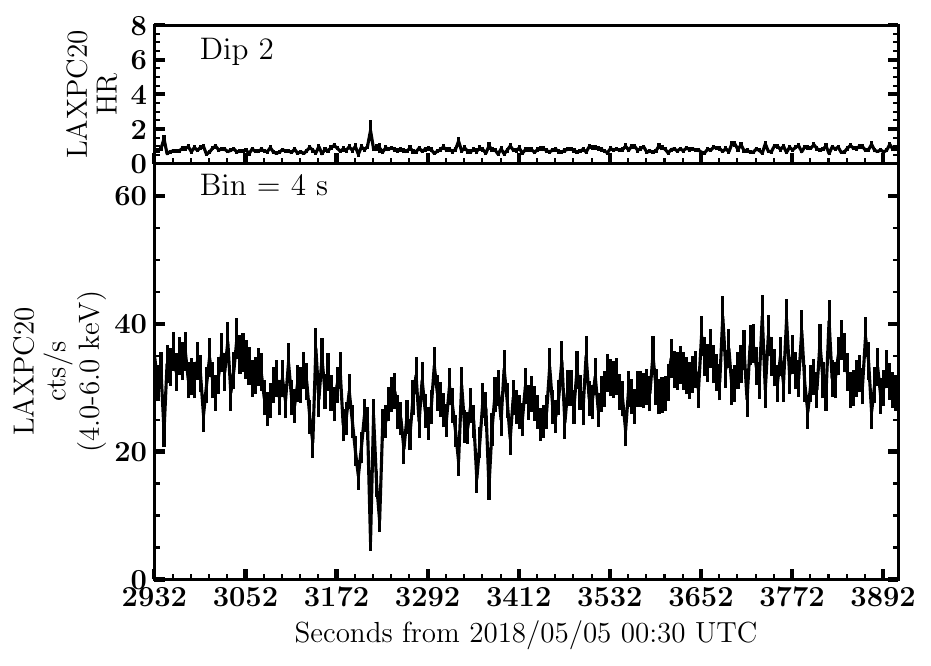}
    \caption{\astrosat{}/LAXPC20 4.0 -- 6.0 keV light curve and hardness ratio (counts between 6.0 -- 20.0 keV divided by counts between 4.0 -- 6.0 keV) with a time bin of 1 s (upper panels) and 4 s (lower panels). The plots on the left side are for dip 1 (shallow dip) and on the right side are for dip 2 (deep dip) observed in \xb.}
    \label{fig:fig4}
\end{figure*}

The LAXPC20 4.0 -- 6.0 keV background-corrected light curve and the hardness ratio (counts between 6.0 -- 20.0 keV energy band divided by the counts between 4.0 -- 6.0 keV energy band) plot of a shallow dip with a time resolution of 1 s is shown in the upper left panel of Figure \ref{fig:fig4}. The part of the same dip with 4 s time resolution is shown in the lower-left panel of the figure. Light curve and hardness ratio for the deep dip are shown in the right panel of Figure \ref{fig:fig4} (upper right panel is with 1 s time resolution and the lower right panel is with 4 s time resolution). If we examine the plot of deep dip with 1 s time resolution, in the beginning, the intensity is decreasing slightly (known as the shoulder of the deep dip). In contrast, the hardness ratio is not varying much, and after some point, both light curve and hardness ratio vary extensively. The strongest hardening corresponds to the deepest segment of the dips. In these figures, we can observe that the shallow dip's depth and hardness ratio is similar to the deep dip, as it comprises many very short duration deep dips.  

\begin{figure*}
    \centering
    \includegraphics[width=0.45\linewidth]{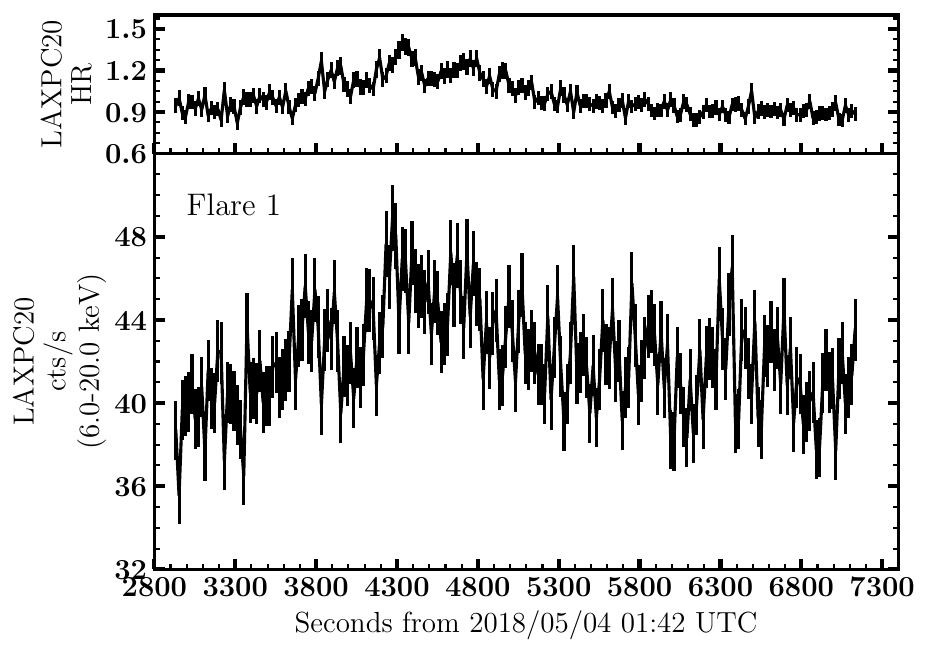}
    \includegraphics[width=0.45\linewidth]{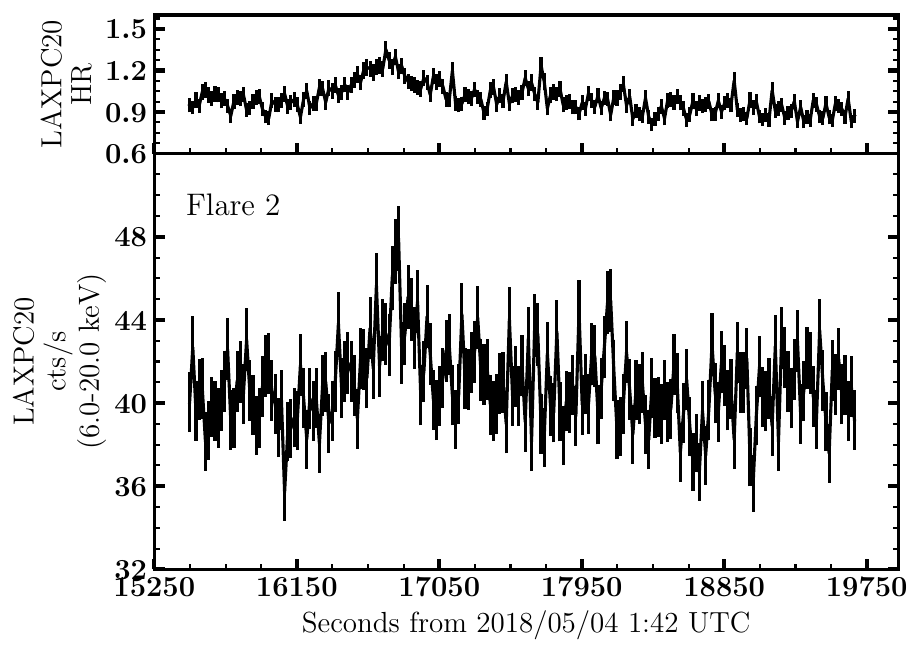}
    \caption{The 6.0 -- 20.0 keV LAXPC20 light curve and the hardness ratio (counts between 6.0 -- 20.0 keV divided by counts between 4.0 -- 6.0 keV) for flare 1 (left panel) and flare 2 (right panel) observed in \xb{} during this observation with a time bin of 20 s.}
    \label{fig:fig5}
\end{figure*}

The two prominent flares observed in \xb{} during this observation are shown in Figure \ref{fig:fig5}. The left panel is flare 1, and the right panel is flare 2. The lower section of each panel is the 6.0 -- 20.0 keV LAXPC20 light curve, and the upper section is the hardness ratio (counts between 6.0 -- 20.0 keV divided by counts between 4.0 -- 6.0 keV). Both light curve and hardness ratio are binned over 20 s.

\subsection{Power Spectral Density (PSD)} \label{subsec:psd}
\begin{figure}
    \centering
    \includegraphics[width=1.0\linewidth]{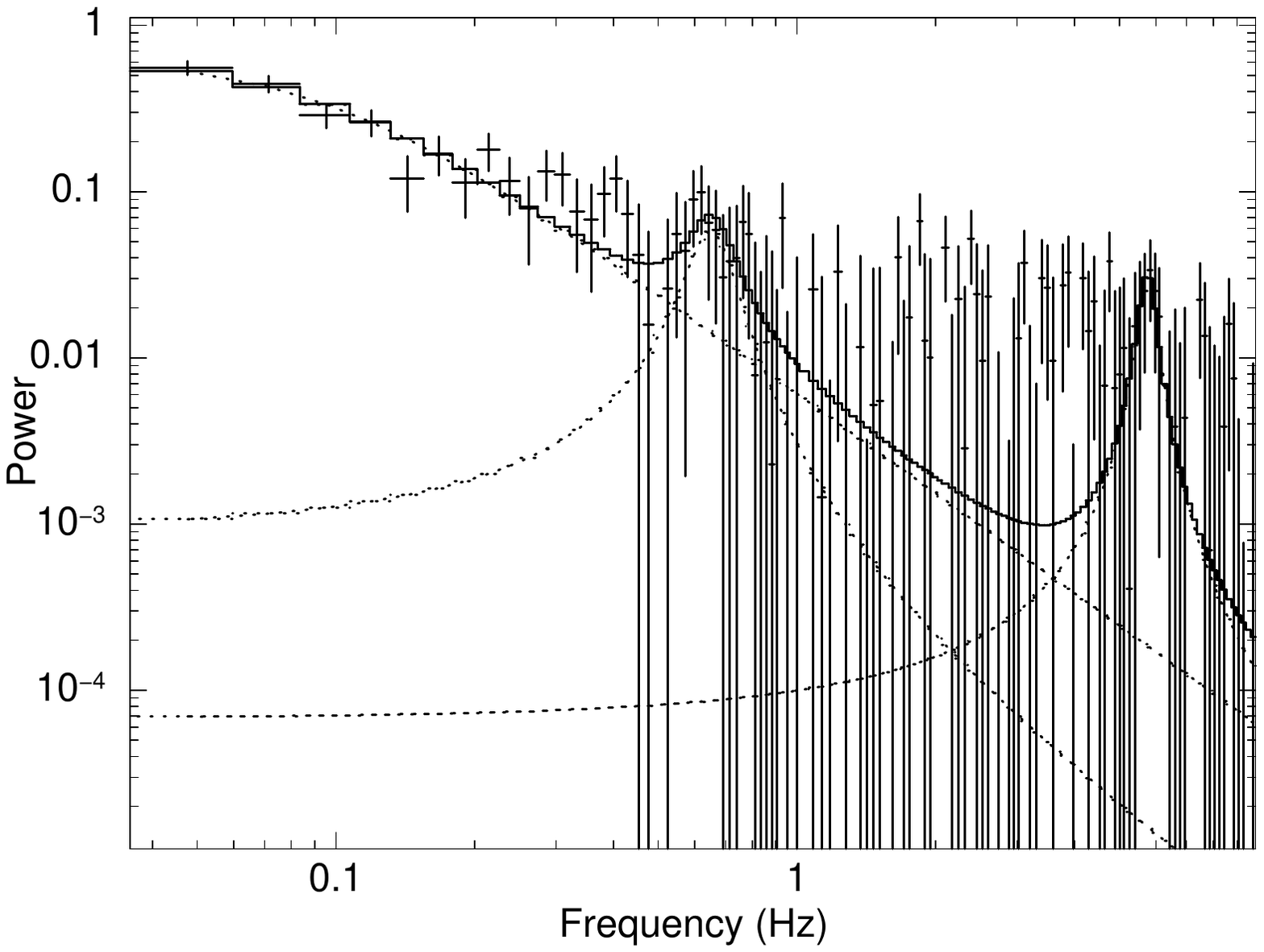}
    \caption{A QPO with significance $\sim$ 2 $\sigma$ and frequency 0.65 Hz seen in \astrosat{}/LAXPC observation of \xb{}. The PDS is extracted using the complete light curve in the 3.0 -- 20.0 keV energy range of LAXPC10 and LAXPC20.}
    \label{fig:fig6}
\end{figure}

\begin{figure}
    \centering
    \includegraphics[width=1.0\linewidth]{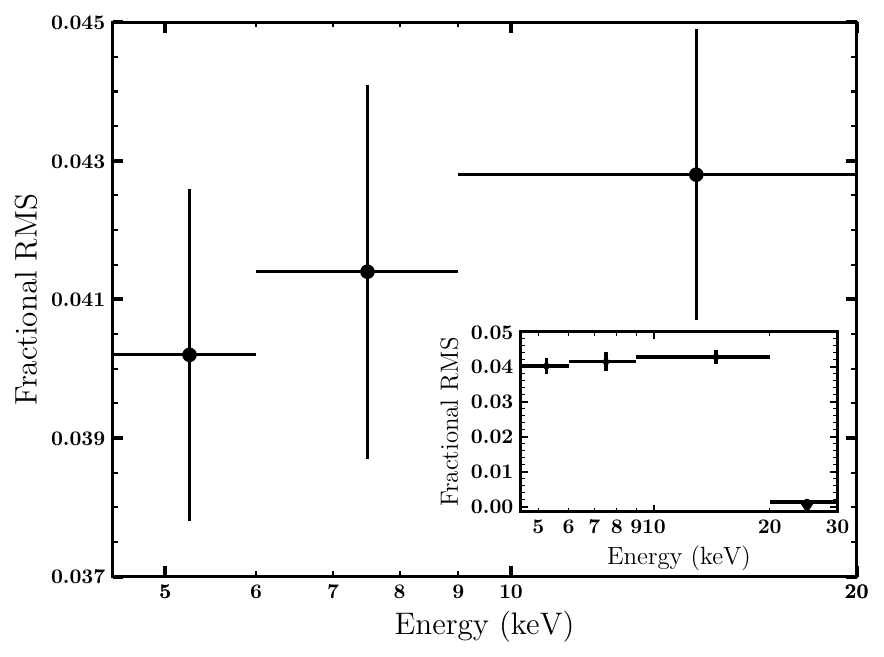}
    \caption{Fractional RMS and energy correlation from the \astrosat{} observation of \xb{}. The RMS was calculated from the PDS of the complete light curve in the 3.0 -- 20.0 keV energy range of LAXPC10 and LAXPC20. The inset shows the relatively low rms estimated in the energy range beyond 20 keV.}
    \label{fig:fig7}
\end{figure}

We used the LAXPC10 and LAXPC20 observations to extract the Power Density Spectra (PDS), using the General High-energy Aperiodic Timing Software ($GHATS$)\footnote{\url{http://astrosat.iucaa.in/~astrosat/GHATS_Package/Home.html}}. We extracted PDS in the energy range 3.0 -- 20.0 keV with Nyquist frequency up to $\sim$3000 Hz using intervals of $\sim$10 s. Inspection of the PDS in various energy bands as well as the sections based on intensity (see Figure \ref{fig:fig6}) does not reveal the presence of high frequency candidate Quasi Periodic Oscillations (QPOs) reported in \cite{2011MNRAS.411.2717M}. We also examined the dynamic PDS during the flares but did not detect any significant high frequency QPOs. To examine the low frequency variability, we extracted a PDS with Nyquist frequency extending up to 10 Hz using $\sim$50 s intervals of the light curve and fit it using a zero centered Lorentzian and a Lorentzian for the QPO. The PDS of the complete light curve in the 3.0 -- 20.0 keV energy range indicates the presence of a QPO with frequency 0.65 Hz (Figure \ref{fig:fig6}) with low statistical significance ($\sim$2 $\sigma$). Examination of the PDS in narrower energy bands indicates the presence of the feature but the statistical significance falls rapidly and there is no conclusive rms--energy relation due to the resulting large uncertainties. A correlation of rms and energy for the QPO can identify it as a 1 Hz QPO usually seen in dipping LMXBs \citep{2012ApJ...760L..30H}. 
Up to 20 keV the rms of the low frequency noise (LFN) and the energy have a Pearson correlation coefficient $\sim 0.59$ (Figure \ref{fig:fig7}). 
Beyond 20.0 keV the rms drops rapidly and we can only estimate upper limits, this is seen from the inset of Figure \ref{fig:fig7}.

\section{Spectral Analysis} \label{subsec:spec_analysis}

The joint SXT and LAXPC20 spectral fitting for each of the four sections of the HID is carried out using \textit{XSPEC v-12.13.1} \citep{1996ASPC..101...17A}. The energy ranges used for SXT and LAXPC20 spectra are between 0.7 -- 7.0 keV and 4.0 -- 20.0 keV, respectively. The counts below 0.7 keV and above 7.0 keV for SXT and above 20.0 keV for LAXPC20 are avoided to minimise the large systematic errors. Also, counts below 4.0 keV for LAXPC20 are ignored due to the uncertainty in the response matrix. For SXT the ARF file \textit{sxt\_pc\_excl00\_v04\_20190608.arf}, RMF file \textit{sxt\_pc\_mat\_g0to12.rmf} and the background file \textit{SkyBkg\_comb\_EL3p5\_Cl\_Rd16p0\_v01.pha} are used. The RMF file \textit{lx20cshp01L1v1.0.rmf} is used for LAXPC20. All these files are obtained from the payload operation center (POC) of SXT\footnote{\url{https://www.tifr.res.in/~astrosat_sxt/dataanalysis.html}} and LAXPC\footnote{\url{https://www.tifr.res.in/~astrosat_laxpc/LaxpcSoft.html}}. The background spectrum for LAXPC20 is generated using \textit{laxpc\_make\_backspectra} tool of \textit{LAXPCSoft}. For the ease of $\chi^2$ fitting in joint spectral analysis, the re-binning of the channels is performed to match the intrinsic spectral resolution of SXT\footnote {Page 52 of \astrosat{} Handbook Ver : 1.11 \url{http://www.iucaa.in/~astrosat/AstroSat_handbook.pdf}} and LAXPC\footnote {Page 6 of \astrosat{} Handbook Ver : 1.11 \url{http://www.iucaa.in/~astrosat/AstroSat_handbook.pdf}}. A systematic uncertainty of 3\% is added while performing the spectral fitting. The SXT data are corrected for the gain; we used \textit{gain fit} command in \textit{XSPEC}, keeping slope fixed to unity. We also included cross-normalisation constant using \textit{constant} model in \textit{XSPEC} for the joint spectral fitting, considering it to be unity for SXT and allowing to vary for the LAXPC20 spectrum. This help us to account for the cross-calibration normalisation uncertainties in the fit. We also excluded dips while performing the spectral fitting of section 1 for the source.  

The spectra from  SXT $\&$ LAXPC20 data are jointly fitted with two sets of models. The first model [\texttt{cons*(tbabs*warmabs*(diskbb+comptt)+ tbabs*(gau$_{1}$+gau$_{2}$))}], called M1 here, is  the one used by \cite{2009A&A...493..145D} to analyse the $XMM\_Newton$ and $INTEGRAL$ data of \xb{}. Some of the parameters in this model, however, could not be constrained and had to be held constant to the values reported in \cite{2009A&A...493..145D}. Hence, we also used a second more generic model [\verb'cons*tbabs*simpl*diskbb'], called M2, to elicit the source behaviour.

\subsection{Flux Resolved Spectroscopy} \label{subsec:fluxrslspec}

\subsubsection{Model M1}
\label{subsubsec:modelM1}
\begin{figure}
    \centering
    \includegraphics[width=1.0\linewidth]{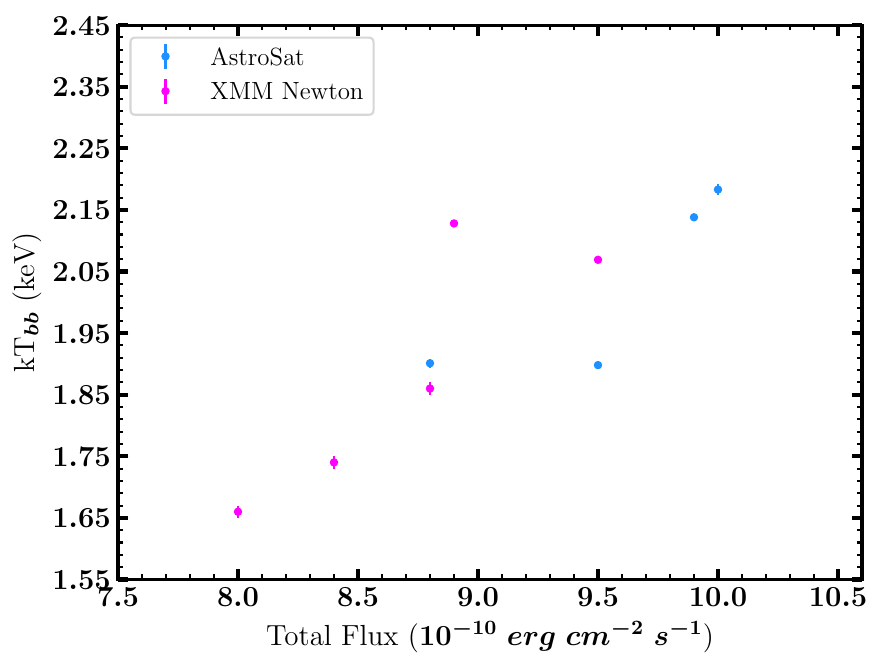}
    \caption{The disc blackbody component from model M1 (see section \ref{subsubsec:modelM1}) as a function of flux. The flux is calculated between 0.6 -- 10.0 keV energy range to match the \textit{XMM-Newton} range. The blue points are \astrosat{} results and the pink points are the \textit{XMM-Newton} and \textit{INTEGRAL} results reported in ~\protect\cite{2009A&A...493..145D}.}
    \label{fig:fig8}
\end{figure}

The spectra are generated using GTIs for each of the HID sections (mentioned in section \ref{subsec:lxp_hid}). We fit the joint SXT+LAXPC20 spectra in the energy range 0.7 -- 20.0 keV for all four intensity dependent sections for \xb{} using the model M1, which is used by \cite{2009A&A...493..145D} to fit \textit{XMM-Newton} and \textit{INTEGRAL} data. The model M1 is consisting of a disc blackbody \verb'(diskbb)' and a thermal comptanization component \verb'(comptt)'. Both these components are modified by neutral photo-electric absorption (\textit{N}$_\mathrm{H}^\mathrm{abs}$) and ionised photo-electric absorption using \verb'warmabs' model, along with two \verb'Gaussian' emission components. The \verb'warmabs' model is consisting of column density of absorber (\textit{N}$_\mathrm{H}^\mathrm{warmabs}$), ionisation parameter $\xi$ = L/nR$^{2}$ (L is the luminosity of the incident radiation, n is the gas density and R is the distance from the radiation source),  turbulent velocity broadening \textit{$\sigma$}$_\mathrm{v}$, and the average systematic velocity shift of the absorber $\nu$. The \verb'warmabs' is used to model the photo-ionised plasma absorption in the line of sight. While fitting the SXT+LAXPC20 spectra we kept the \verb'warmabs' and both \verb'Gaussian' component parameters fixed to that of the \textit{XMM-Newton} and \textit{INTEGRAL} parameters reported in \cite{2009A&A...493..145D}. Also to make a reasonable comparison between our results and \textit{XMM-Newton} and \textit{INTEGRAL} results, we calculated the flux between 0.6 -- 10.0 keV range by standardising the energy range using the \textit{energies} command within \textit{XSPEC} while fitting.    
A significant change in the disc blackbody component is observed when we fitted the spectra by keeping \verb'warmabs' and \verb'Gaussian' parameters consistent with the \textit{XMM-Newton} and \textit{INTEGRAL} results. The detailed best fit spectral parameters using M1 for HID sections 1 to 4 (sections given in Figure \ref{fig:fig2}) are given in Table \ref{tab:paramM1}. The disc blackbody temperature (\textit{kT}$_\mathrm{bb}$) as a function of flux is shown in Figure \ref{fig:fig8}, along with the \textit{XMM-Newton} and \textit{INTEGRAL} values reported in Table 4 of \cite{2009A&A...493..145D}. We found that the change in the disc blackbody temperature is consistent with that of the \textit{XMM-Newton} and \textit{INTEGRAL} values. We observe the shift in the flux values for \astrosat{} results as we are considering the entire observation for our study without excluding flares; though, we are not considering the dips while fitting the spectra. Note that \cite{2009A&A...493..145D} have considered only persistent emission from all 5 observations (2 observations with deep dips, 1 with shallow dip and 2 without dips) excluding dips and bursts.

\begin{table*}
\small
\centering
\caption{ Spectral fitting parameters of \xb{} spectra in the banana state observation with \astrosat{}. Joint SXT+LAXPC20 spectra for four sections are fitted with model M1:  \texttt{cons*(tbabs*warmabs*(diskbb+comptt)+tbabs*(gau$_{1}$+gau$_{2}$))}. Errors represent 90\% confidence intervals.}
\label{tab:paramM1}
\begin{tabular}{llccccc}
\hline\\
& & \multicolumn{4}{c}{\textit{AstroSat}} & \textit{XMM Newton \&}   \\ 
\\
HID Section & & 1 & 2 & 3 & 4 & \textit{INTEGRAL}$^\dag$ \\
\\
\hline\\
& Comp. & & & & \\
Parameter & & & & \\

& \verb'comptt' & & & & \\
\textit{kT}$_\mathrm{0}$ (keV) & & 0.222 $\pm$ 0.008  & 0.261 $\pm$ 0.015 & 0.216 $\pm$ 0.015 & 0.231 $\pm$ 0.013 & 0.173 $\pm$ 0.001 \\

\textit{kT}$_\mathrm{comptt}$ (keV) & & 4.61 $\pm$ 0.06 & 3.34 $\pm$ 0.03 & 3.65 $\pm$ 0.04 & 3.28 $\pm$ 0.03 & 3.06 $\pm$ 0.03 \\

$\tau_\mathrm{p}$ & & 4.42 $\pm$ 0.06 & 7.44 $\pm$ 0.10 & 7.09 $\pm$0.11 & 7.90 $\pm$ 0.10 & 6.07 $\pm$ 0.05 \\

\textit{k}$_\mathrm{comptt}$ & & 0.0287 $\pm$ 0.0008   & 0.0191 $\pm$ 0.0006  & 0.0165 $\pm$ 0.0007  & 0.0325 $\pm$ 0.0009  & 0.0763$_{-0.0003}^{+0.0006}$  \\

& \verb'dbb' & & & & \\
\textit{kT}$_\mathrm{bb}$ (keV) & & 1.901 $\pm$ 0.007 & 1.898 $\pm$ 0.005  & 2.138 $\pm$ 0.006 & 2.183 $\pm$ 0.009 & 1.86 $\pm$ 0.01 \\

\textit{k}$_\mathrm{bb}$ [(R$_\mathrm{in}$/D$_\mathrm{10}$)$^{2}$ cos$\theta$] & & 2.78 $\pm$ 0.04 & 3.32 $\pm$ 0.04 & 2.29 $\pm$ 0.02 & 1.78 $\pm$ 0.03 & 1.76$_{-0.03}^{+0.01}$\\

& \verb'gau'$_{1}$ &  &  &  &  \\
\textit{E}$_\mathrm{gau}$ (keV) & & 6.8  $\pm$ 0.3 & 7.0 $\pm$ 0.2 & 6.9 $\pm$ 0.3 & 7.0 $\pm$ 0.2 & 6.59 $\pm$ 0.05 \\

\textit{$\sigma$} (keV) & & 0.53$^f$ & 00.53$^f$ & 0.53$^f$ & 0.53$^f$ & 0.53 $\pm$ 0.06 \\

$k_\mathrm{gau}$ (10$^{-4}$ ph cm$^{-2}$ s$^{-1}$) & & 2.0 $\pm$ 0.5 & 2.2 $\pm$ 0.5 & 2.3 $\pm$ 0.6 &  2.9 $\pm$ 0.7 & 3.3 $\pm$ 0.3 \\

& \verb'gau'$_{2}$ & & & & \\
\textit{E}$_\mathrm{gau}$ (keV) & & 0.85 $\pm$ 0.04 & 0.92 $\pm$ 0.09 & <1.73 & 0.84$_{-0.10}^{+0.047}$ & 1.046 $\pm$ 0.004 \\

\textit{$\sigma$} (keV) & & 0.1$^f$ & 0.1$^f$ & 0.1$^f$ & 0.1$^f$ & 0.1 \\

\textit{k}$_\mathrm{gau}$ (10$^{-3}$ ph cm$^{-2}$ s$^{-1}$) & & 1.2$_{-0.2}^{+0.3}$ & 2.3 $\pm$ 0.1 & <0.67 & 3.9 $\pm$ 0.2 & 1.69$_{-0.07}^{+0.04}$   \\

& \verb'tbabs' & & & & \\
\textit{N}$_\mathrm{H}^\mathrm{abs}$ (10$^{22}$ cm$^{-2}$) & & 0.20 $\pm$ 0.01 & 0.19 $\pm$ 0.01 & 0.15 $\pm$ 0.01 & 0.22 $\pm$ 0.01 & 0.220 $\pm$ 0.001 \\

& \verb'warmabs' & & & & \\
\textit{N}$_\mathrm{H}^\mathrm{warmabs}$ (10$^{22}$ cm$^{-2}$) & & 2.0$^f$ & 2.0$^f$ & 2.0$^f$ & 2.0$^f$ & 2.0 $\pm$ 0.3 \\

log\textit{($\xi$)} (erg cm s$^{-1}$) & & 3.95$^f$ & 3.95$^f$ & 3.95$^f$ & 3.95$^f$ & 3.95 \\

\textit{$\sigma$}$_\mathrm{v}$ (km s$^{-1}$) & & 2600$^f$ & 2600$^f$ & 2600$^f$ & 2600$^f$ & 2600$_{-1300}^{+1700}$  \\

\textit{$\nu$} (km s$^{-1}$) & & & & & & -1400 $\pm$ 780 \\
\\

\hline\\
\textit{$\chi_{v}^{2}$} (d.o.f.) & & 0.99 (53) & 1.13 (56) & 0.90 (56) & 1.09 (56) & 1.51 (1147) \\
\\
\hline\\
\textit{F} ($10^{-10}$ erg $cm^{-2}$ $s^{-1}$) & & 8.8 & 9.5 & 9.9 & 10.0 & 8.8 \\
(0.6 -- 10.0 keV) & & & & & &  \\
[1ex]

\textit{F}$_\mathrm{dbb}$ ($10^{-10}$ erg $cm^{-2}$ $s^{-1}$) & & 7.1 & 8.0 & 9.1 & 8.6 & 4.1 \\
(0.6 -- 10.0 keV) & & & & & &  \\
[1ex]

Exposure (ks) & & 6.1$^s$/19.1$^l$ & 12.9$^s$/33.6$^l$ & 11.3$^s$/35.6$^l$ & 3.5$^s$/7.2$^l$ & 26.6 \\
\\
\hline
& \verb'True Inner Disc Radius' & & & & & \\
\textit{r}$_\mathrm{in}$ (km)$^\dag$$^\dag$ & & 6.12 $\pm$ 0.04 & 6.69 $\pm$ 0.04 & 5.55 $\pm$ 0.02 & 4.90 $\pm$ 0.03 & \\
\hline\\
\end{tabular}
\begin{flushleft}
 
 \textbf{Note:}
$^f$ represents that the parameters are fixed to the values reported in \cite{2009A&A...493..145D} while fitting SXT \& LAXPC20 joint spectra. $^s$ and $^l$ denotes the exposure time for SXT and LAXPC20 corresponding to each HID section. Note $^\dag$$^\dag$f = 1.7, color or correction factor. \\    
$^\dag$\cite{2009A&A...493..145D}
\end{flushleft}
\end{table*}

\subsubsection{Model M2}
\label{subsubsec:modelM2}
\begin{figure}
    \centering
    \includegraphics[width=1.0\linewidth]{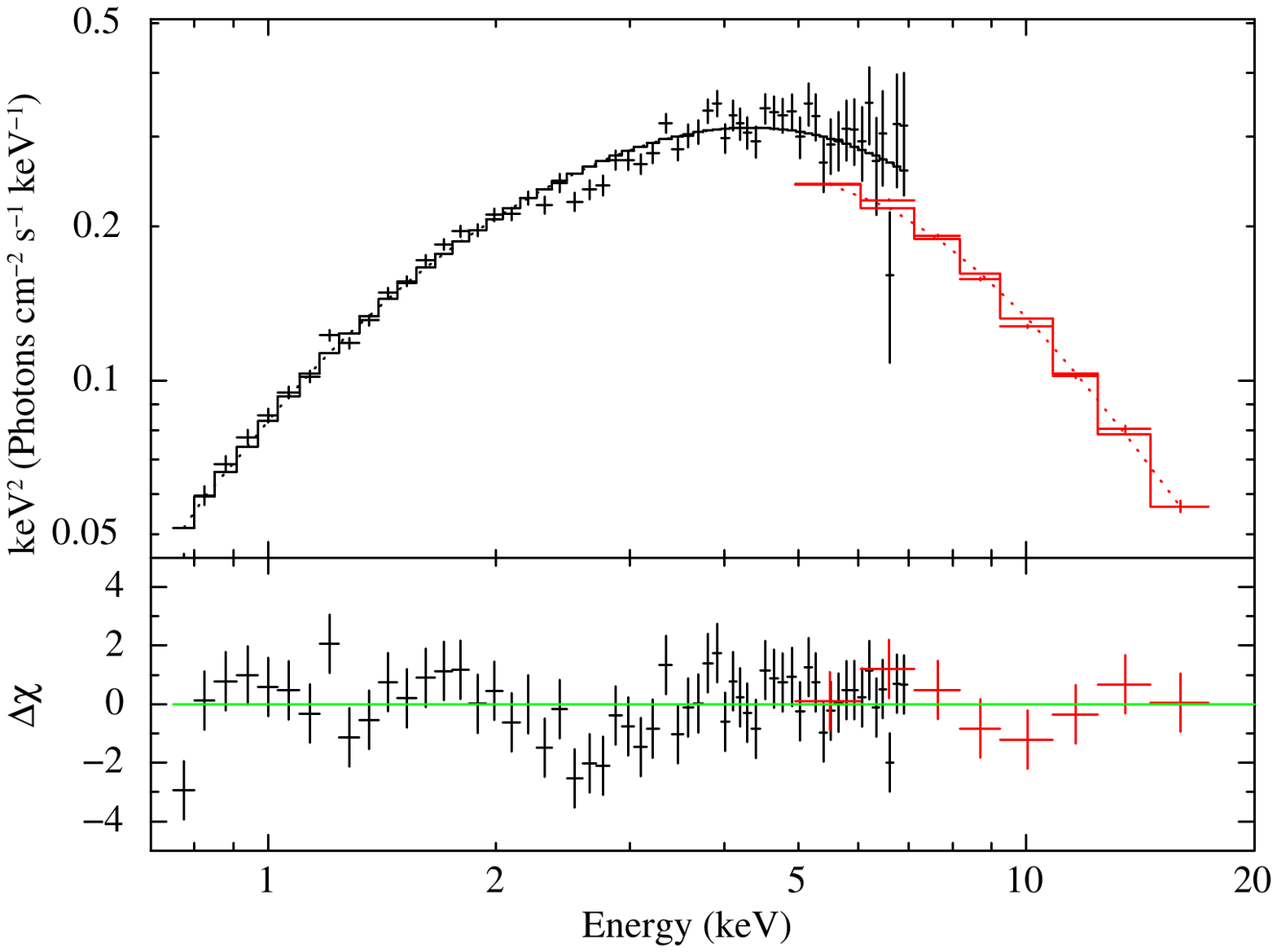}
    \caption{The results of 0.7 -- 20.0 keV \astrosat{} spectral fitting of \xb{} for the first section of the HID (see section \ref{subsec:lxp_hid} for details on HID sections). Top panel: the SXT+LAXPC20 spectra. Black points indicate SXT (0.7 -- 7.0 keV) data and red represents the LAXPC20 (4.0 -- 20.0 keV) data. The spectra are fitted with model M2: \texttt{cons*tbabs*cflux*simpl*diskbb}. Bottom panel: Residuals from the model fit.}
    \label{fig:fig9}
\end{figure}

The joint spectra of SXT \& LAXPC20 in the energy range of 0.7 -- 20.0 keV are fitted with the model M2 consisting of a combination of Comptonization (\verb'simpl': \cite{2009PASP..121.1279S}) and a multicoloured disc blackbody (\verb'diskbb': \cite{Mitsuda_1984}, \cite{Makishima_1986}) along with Galactic absorption (\verb'TBabs'). We implemented the convolution model \verb'cflux' in this model, to estimate the total flux and disc flux for the energy range 0.7 -- 20.0 keV and calculated the coronal flux of the source (Figure \ref{fig:fig12}). The best-fit spectra for the first HID section (section 1 (black) given in Figure \ref{fig:fig2}) are shown in Figure \ref{fig:fig9}, along with the residuals. The detailed best-fit spectral parameters for HID sections 1 to 4 are given in Table \ref{tab:paramM2}. 

\begin{figure}
    \centering
    \includegraphics[width=1.0\linewidth]{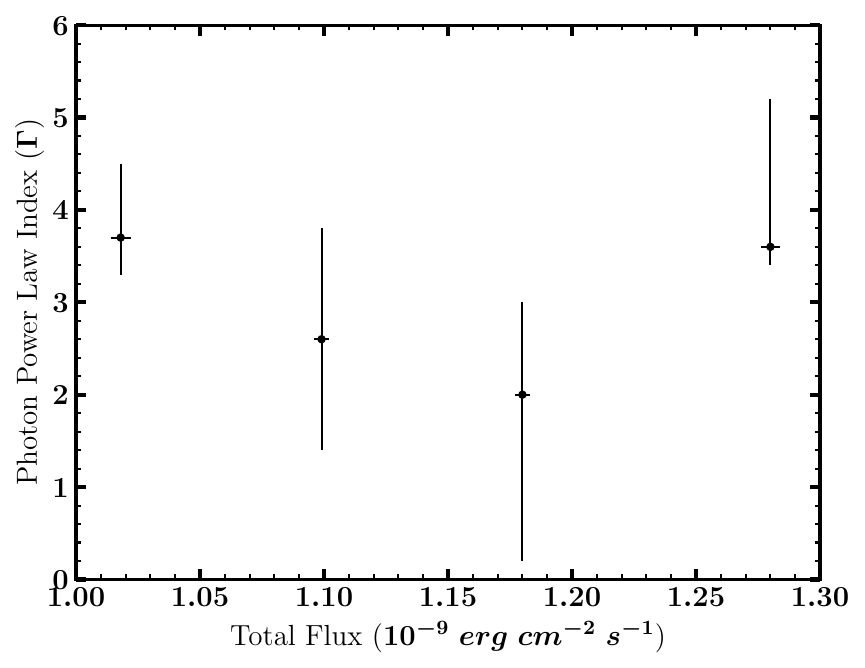}
    \caption{The results of 0.7 -- 20.0 keV \astrosat{} spectral fitting of \xb{} for the four HID sections (see section \ref{subsec:lxp_hid} for details on HID sections). Spectral Fitting parameter: Photon Power Law Index ($\Gamma$) as a function of Total Flux. We do not observe a notable trend within the uncertainties.}
    \label{fig:fig10}
\end{figure}

\begin{figure}
    \centering
    \includegraphics[width=1.0\linewidth]{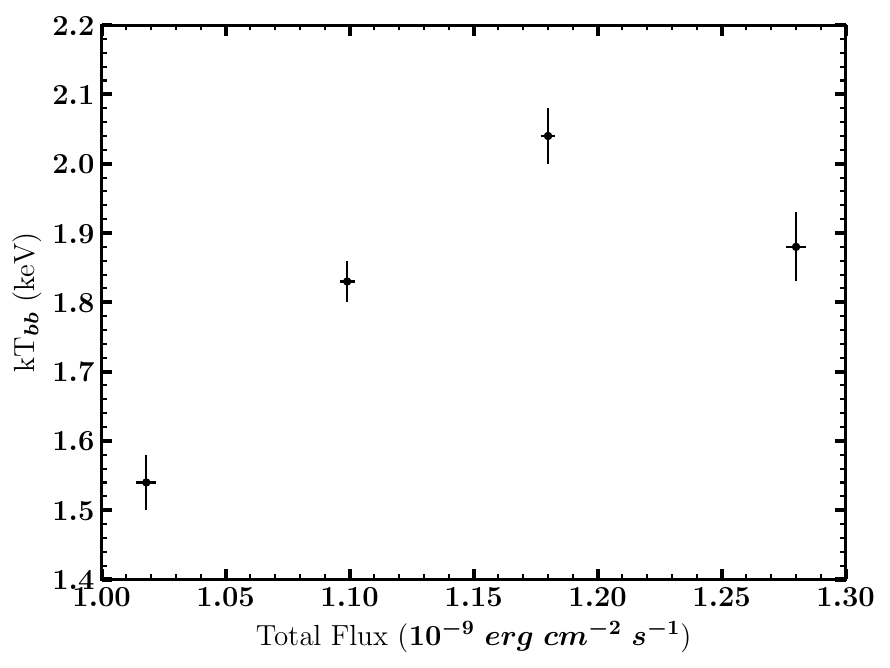}
    \caption{The results of 0.7 -- 20.0 keV \astrosat{} spectral fitting of \xb{} for the four HID sections (see section \ref{subsec:lxp_hid} for details on HID sections). Spectral Fitting parameter: Diskbb Temperature (\textit{kT}$_\mathrm{in}$) as a function of Total Flux, found to be increasing as the source become harder.}
    \label{fig:fig11}
\end{figure}

\begin{figure}
    \centering
    \includegraphics[width=1.0\linewidth]{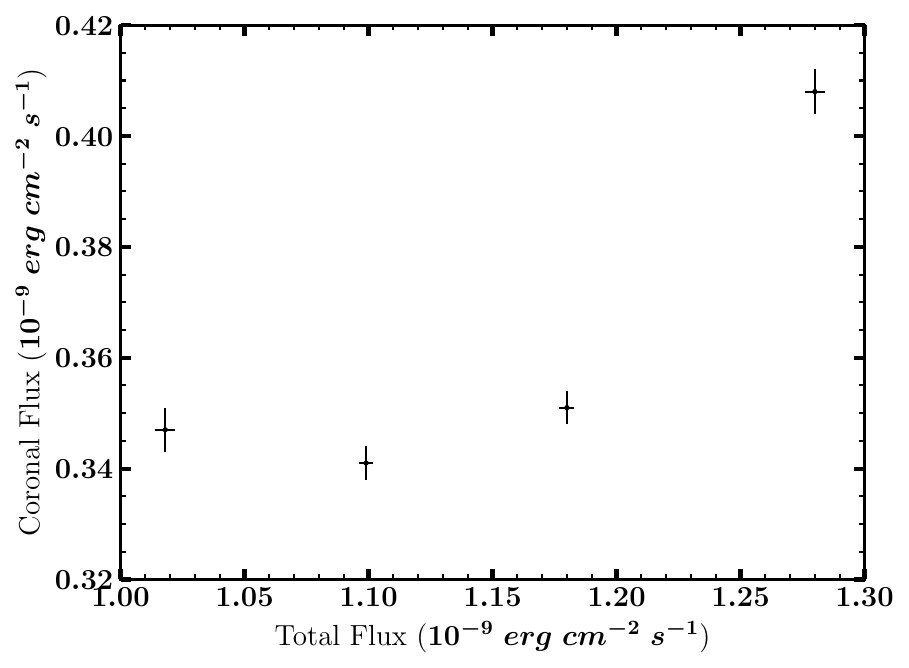}
    \caption{Coronal flux of \xb{} as a function of total Flux for the four sections of HID (see section \ref{subsec:lxp_hid} for details on HID sections).}
    \label{fig:fig12}
\end{figure}

We observed the variations in the inner disc temperature (\textit{kT}$_\mathrm{bb}$) as a function of total flux (Figure \ref{fig:fig11}). We did not observe significant variations in the photon power-law index ($\Gamma$) (Figure \ref{fig:fig10}). The best-fit offset values for the SXT gain fit were found to be 42.41 eV, 34.26 eV, 53.88 eV and 44.33 eV for sections 1, 2, 3 and 4, respectively. Furthermore, we did not observe improvement in the fitting parameter values by adding an iron line to the spectral model. 

\begin{table*}
\centering
\caption{Spectral fitting parameters of \xb{} spectra in the banana state observation with \astrosat{}. Joint SXT+LAXPC20 spectra is fitted with model M2:  \texttt{cons*tbabs*cflux*simpl*diskbb}. Errors represent 90\% confidence intervals. True Inner Disc Radius for \xb{} calculated using disc blackbody normalization for each HID section}
\label{tab:paramM2}
\begin{tabular}{llcccc}
\hline\\
& & \multicolumn{4}{c}{\textit{AstroSat}}  \\ 
\\
HID Section & & 1 & 2 & 3 & 4 \\
\\
\hline\\
& Comp. & & & & \\
Parameter & & & & \\
[1ex]
\\
& \verb'tbabs' & & & & \\
\textit{N}$_\mathrm{H}^\mathrm{abs}$ (10$^{22}$ cm$^{-2}$) & & 0.08 $\pm$ 0.02 & 0.10 $\pm$ 0.01 & 0.08 $\pm$ 0.01 & 0.09 $\pm$ 0.01 \\
[1ex]
\\
& \verb'Total Flux' & & & & \\
& (0.7 -- 20.0 keV) & & & \\
\verb'cflux' (10$^\mathrm{-9}$ erg cm$^\mathrm{-2}$ s$^\mathrm{-1}$) & & 1.018 $\pm$ 0.004 & 1.099 $\pm$ 0.003 & 1.180 $\pm$ 0.003 & 1.280 $\pm$ 0.004  \\
[1ex]
\\
& \verb'simpl' & & & & \\
$\Gamma$ & &  3.7$_{-0.8}^{+0.4}$ & 2.6 $\pm$ 1.2 & 2.0$_{-1.0}^{+1.8}$ & 3.6$_{-1.6}^{+0.2}$\\
[1ex]
\verb'FracSctr' & &  0.46$_{-0.06}^{+0.07}$ & 0.17 $\pm$ 0.02 & 0.10 $\pm$ 0.01 & 0.61$_{-0.14}^{+0.19}$\\
[1ex]
\\
& \verb'dbb' & & & &  \\
\textit{kT}$_\mathrm{bb}$ (keV) & & 1.54 $\pm$ 0.04 & 1.83 $\pm$ 0.03 & 2.04 $\pm$ 0.04 & 1.88 $\pm$ 0.05  \\
[1ex]
\textit{k}$_\mathrm{bb}$ [(R$_{in}$/D$_{10}$)$^{2}$ cos$\theta$] & & 7.3$_{-2.3}^{+1.9}$ & 4.4 $\pm$ 0.7 & 3.1$_{-1.0}^{+0.7}$ & 3.8$_{-0.8}^{+0.5}$ \\
[1ex]
& \verb'dbb Flux' & & & & \\
& (0.7 -- 20.0 keV) & & & \\
\verb'cflux' (10$^\mathrm{-9}$ erg cm$^\mathrm{-2}$ s$^\mathrm{-1}$) & & 0.671 $\pm$ 0.005 & 0.758 $\pm$ 0.001 & 0.829 $\pm$ 0.001 & 0.872 $\pm$ 0.001  \\
[1ex]
& \verb'Coronal Flux' & & & & \\
& (0.7 -- 20.0 keV) & & & \\
\verb'cflux' (10$^\mathrm{-9}$ erg cm$^\mathrm{-2}$ s$^\mathrm{-1}$) & & 0.347 $\pm$ 0.004 & 0.341 $\pm$ 0.003 & 0.351 $\pm$ 0.003 & 0.408 $\pm$ 0.004  \\
[1ex]
\hline
\textit{$\chi_{v}^{2}$} (d.o.f.) & & 1.16 (60) & 1.15 (58) & 1.16 (60) & 1.13 (60) \\
 [1ex]
\hline
& \verb'True Inner Disc Radius' & & & & \\
\textit{r}$_\mathrm{in}$ (km)$^\dag$ & & 9.91 $\pm$ 0.02 & 7.66 $\pm$ 0.01 & 6.42 $\pm$ 0.01 & 7.26 $\pm$ 0.02 \\
\hline\\
\end{tabular}
\begin{flushleft}
Note $^\dag$f = 1.7, color or correction factor. Parameter 3: $UpScOnly$ of the \verb'simpl' model is frozen to unity during spectral fitting. 
\end{flushleft}
\end{table*}

\subsection{Constraining Disc Inner Edge Radius} \label{subsec:inneredgerad}

\begin{figure}
    \centering
    \includegraphics[width=1.0\linewidth]{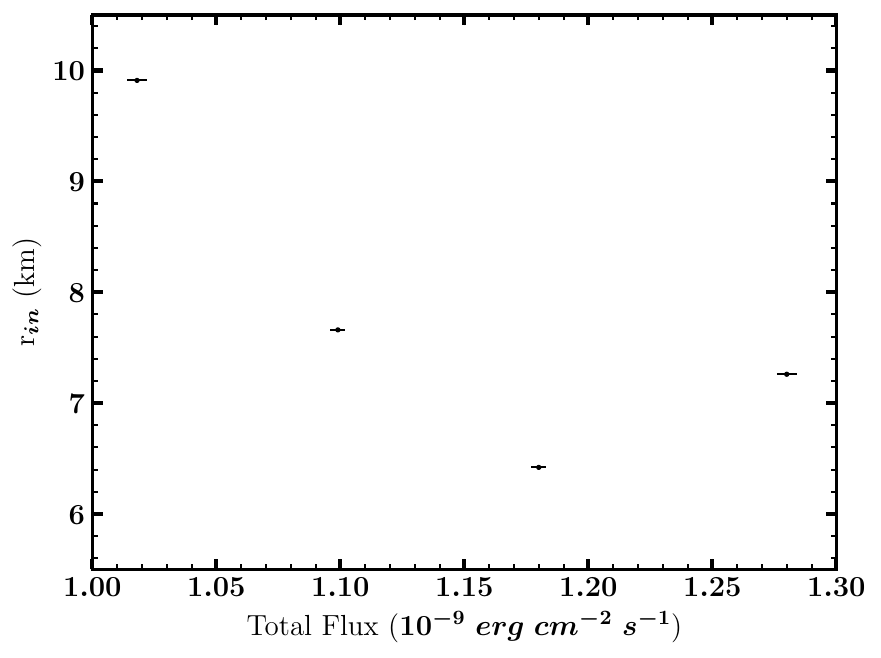}
    \caption{True Inner disc Radius (\textit{r}$_\mathrm{in}$) in km for \xb{} as function of total flux for each HID section, calculated using \texttt{diskbb} normalization \textit{k}$_\mathrm{bb}$.}
    \label{fig:fig13}
\end{figure}

The normalization parameter of \verb'diskbb' model is defined as (R$_\mathrm{in}$/D$_\mathrm{10}$)$^\mathrm2$ $\cos\theta$, where R$_\mathrm{in}$ is the apparent inner disc radius in km, D$_\mathrm{10}$ is the distance to the source in 10 kpc units, and $\theta$ is the angle of the disc with respect to the observer. In our spectral fitting using M1 and M2, we observed the variation in the diskbb normalization (\textit{k}$_\mathrm{bb}$) for section 1, section 2, section 3 and section 4 of HID. We calculated the Disc Inner Edge Radius (\textit{r}$_\mathrm{in}$) using diskbb normalization from both M1 and M2 (Figure \ref{fig:fig13}). We considered distance to source as 7.6 $\pm$ 0.8 kpc \citep{2017RAA....17..108G}, inclination angle between 65$^{\circ}$ -- 73$^{\circ}$ \citep{1987ApJ...313..792M,2009A&A...493..145D} and a color or a correction factor ($f$) as 1.7 \citep{1995ApJ...445..780S}. The Disc Inner Edge Radius (\textit{r}$_\mathrm{in}$) is found to be decreasing as the source became harder from section 1 to section 4 (Figure \ref{fig:fig13}). The calculated values of \textit{r}$_\mathrm{in}$ in km for each section are given in Table \ref{tab:paramM2}.

\section{Discussion} \label{sec:discussion}

In this paper, we report the results of the flux resolved spectroscopy of \xb{} data observed with \astrosat{} between 2018 May 03 and 2018 May 05. We have utilised the data from two pointing instruments (SXT and LAXPC) onboard \astrosat{}. The hardness ratio of the source is variable. \xb{} is found to be exhibiting flares at the beginning of this observation. The flaring phenomenon is a high energy phenomenon dominated by hard X-rays and hence were found to be associated with the hardness ratio and intensity. We observe two dips after the flaring episodes during this observation. The first dip we observed is a ``shallow'' dip, and the second is a ``deep'' dip. Dips are related to an increased absorption at low energies and, therefore, with an anti-correlation between hardness and intensity. The first dip with 100 s time resolution is identified as ``shallow'', is consisting of a few rapid, deep variations with a time resolution of 1 s. The characteristics of dips are consistent with that observed in the earlier observations \citep{2009A&A...493..145D}. The presence of flares, along with dips and thermonuclear X-ray bursts, has been reported earlier \citep{2002ApJ...581.1286S,2011MNRAS.411.2717M}. This is the first time we observed strong flares in \xb{} at the beginning, which is followed by two dips. The X-ray dipping properties of the source have been studied extensively in earlier observations \citep{2001ApJ...548..883I,2002ApJ...581.1286S,2007A&A...464..291I,2011MNRAS.411.2717M}. Most of the studies have reported the properties of persistent emission alone. Nevertheless, a detailed study of \xb{} data inclusive of flares (and excluding dips) have not been done earlier.

\xb{} was found to be in a banana state during this observation. \cite{2007MNRAS.377..198B} established that \xb{} is an atoll source. In the island state, an atoll source does not move up and down so fast, and the track is not as well-defined as in the banana state (see review by \textcolor{blue}{Lewin \& van der Klis} ({\textcolor{blue}{2006})). The shape of the track in the HID, after comparing with the HID tracks of atoll sources (e.g., in \textcolor{blue}{Lewin \& van der Klis} ({\textcolor{blue}{2006})), including the HID track of \xb{} given in \cite{2007MNRAS.377..198B}, strongly indicates that the source is in the banana state during this observation.

We have performed a flux resolved spectroscopy on the spectra extracted at the different intensity levels with changing hardness. \cite{2011MNRAS.411.2717M} reported the systematic continuum (excluding flares, dips, and bursts) spectral analysis of \xb{} using $RXTE$ observations. They performed analysis using 19 various combinations of continuum spectral models (see Table 1 of \cite{2011MNRAS.411.2717M}) on the upper banana state and lower banana state of the source and suggested that the \verb'bknpower'+\verb'comptt' is the preferred model for the continuum. In our study, we considered the observation, which includes flares but excludes dips (note that thermonuclear bursts were not observed) and extracted the spectra from four different sections of HID. The joint SXT and LAXPC20 spectra with total energy of 0.7 to 20.0 keV for each of the HID sections were well fitted with two sets of models---M1: \texttt{cons*tbabs*warmabs*(diskbb+comptt)+tbabs*(gau$_{1}$+gau$_{2}$)} and M2: \texttt{cons*tbabs*cflux*simpl*diskbb}. Model M1  consists of a disc blackbody (\verb'diskbb') and thermal comptanization component (\verb'comptt'), along with neutral photo-electric absorption (\textit{N}$_\mathrm{H}^\mathrm{abs}$) and ionised photo-electric absorption using \verb'warmabs' model and two \verb'Gaussian' emission components. Model M2 consists of Comptonization model and a multicoloured disc blackbody model along with Galactic absorption. In M2, we tried replacing \verb'diskbb' model with the blackbody model, which resulted in a much higher reduced $\chi^{2}$ ($\sim$3.4). Therefore, we suggest that \verb'simpl'*\verb'diskbb' is a non-complex model, suitable for the spectra (which includes flares) of \xb{} in 0.7 -- 20.0 keV energy range.

In model M1 (see Table \ref{tab:paramM1}), we have kept the \verb'warmabs' parameters identical with \textit{XMM-Newton} parameters \citep{2009A&A...493..145D}, to measure the disc blackbody temperature (\textit{kT}$_\mathrm{bb}$) and flux using the Soft X-ray Telescope (SXT). This allowed us to discover the \textit{kT}$_\mathrm{bb}$ and flux correlation for \xb{} for the first time. The disc blackbody temperature is lower for section 1 (1.883), increased gradually for the sections that do not have dips or flares (2.005 and 2.134), and highest for the section which has flares (2.386). The change in the disc blackbody temperature we observed is consistent with the trend observed in the \cite{2009A&A...493..145D}. The significant increase in the blackbody temperature during the flares has been observed in the Z-source Sco X--1 \citep{1985ApJ...296..475W} and the systematic increase in the inner disc temperature of the big dipper $X 1624-490$ has also been previously reported \citep{2001A&A...378..847B}. 

In model M2  (see Table \ref{tab:paramM2}), the lower energy coverage of SXT allowed us to constrain Galactic column density for the source. The photon power-law index as a function of total flux (Figure \ref{fig:fig10}) does not show a significant trend given the uncertainties. In M2, the inner disc temperature (\textit{kT}$_\mathrm{bb}$) is found to increase from Section 1 to Section 3 (Figure \ref{fig:fig11}) with a minimum value of 1.54 $\pm$ 0.04 and a maximum value of 2.04 $\pm$ 0.04. However, it decreases slightly for section 4. We computed the Coronal flux for \xb{} and plotted it as a function of the total flux (Figure \ref{fig:fig12}). The Coronal flux was found to be increasing as the total source flux increased. The disc inner edge radius (r$_{in}$) (Figure \ref{fig:fig13}) calculated using the values of diskbb normalization (\textit{k}$_\mathrm{bb}$) and are given in Table \ref{tab:paramM1} and Table \ref{tab:paramM2}.

Although \xb{} has been the target of many X-ray missions since its discovery in 1984, timing characteristics were not reported until 2011, in the \textit{RXTE} observations of January 6 to March 13, 2008 \citep{2011MNRAS.411.2717M}. In these observations \cite{2011MNRAS.411.2717M} detected high-frequency QPOs $\sim$96 Hz. In the \astrosat{} observations we did not detect the $\sim$96 Hz QPO or the high frequency QPOs reported by \cite{2011MNRAS.411.2717M}. This is probably because the source was in the high-intensity upper banana state (UB), which is also indicated by the very low-frequency noise (VLFN) detected in the PDS \citep{2004astro.ph.10551V, 2006csxs.book...39V}. The RMS of the VLFN tends to increase with energy as seen in Figure \ref{fig:fig7} where we show the relation up to the energy range 20 keV. Beyond 20 keV, the RMS drops as shown in the inset of Figure \ref{fig:fig7}, indicating that the VLFN up to 20 keV is related to the unstable accretion dynamics of matter falling onto the neutron star.
The presence of VLFN ($\sim$3 Hz) also indicates that the source was in the UB during this observation \citep{2004astro.ph.10551V,2006csxs.book...39V,2007MNRAS.377..198B}. LFQPOs, called 1 Hz QPOs, are usually seen in dipping LMXBs like \xb{}. These LFQPOs have not been reported so far in \xb{} \citep{2012ApJ...760L..30H, 2007MNRAS.377..198B, 2011MNRAS.411.2717M}. In our analysis of the \astrosat{} observation, we detect a $\sim$0.65 Hz signal with low significance ($\sim$ 2.7 $\sigma$) that can be confirmed with longer observations.

In summary, for the first time we have found a \textit{kT}$_\mathrm{bb}$ and flux correlation for NS-LMXB \xb{}. We have also given a constraint on the disc inner edge radius and tried to understand its evolution during flares using SXT and LAXPC on board \astrosat{} data through the flux-resolved spectroscopy. Future observation of the source with \astrosat{}, simultaneously with NuSTAR (for the hard band) and NICER (for the soft band), can be used to study the source in detail and to give better constraints to the parameters. Recurrent monitoring of the source can provide a distinctive opportunity to study the evolution of these parameters, and it will help us correlate them.

\section*{Acknowledgements}

NRN and DP acknowledge the financial support from the Indian Space Research Organisation (ISRO) under the \astrosat{} archival Data utilization program. 
This publication uses data from the \astrosat{} mission of the Indian Space Research Organization (ISRO), archived at the Indian Space Science Data Center (ISSDC). 
The authors acknowledge the LAXPC and SXT Payload Operations Centers (POCs) for providing us with essential input and the necessary tools for data analysis.
This research has made use of software provided by the High Energy Astrophysics Science Archive Research Center (HEASARC), which is a service of the Astrophysics Science Division at NASA/GSFC.
This work has made use of the GHATS package developed by Late Prof. Tomaso Belloni at INAF-OAB, Merate, Italy. NRN and DP are thankful to the Department of Physics, Ramniranjan Jhunjhunwala College and University of Mumbai, for providing the necessary facilities to carry out the work. NRN and DP are also grateful to the Tata Institute of Fundamental Research (TIFR), Inter University Center for Astronomy and Astrophysics (IUCAA), and \astrosat{} Science Support Cell (ASSC) hosted by IUCAA and TIFR for their support in the work. NRN thank 
Dr. Sunil Chandra for important suggestions regarding the $SXT$ data normalisation method, Dr. Navin Sridhar for valuable suggestions in the data grouping technique and Dr. Yash Bhargava for valuable comments regarding $SXT$ and $LAXPC$ spectral fitting. The authors thank the referees for providing their suggestions and guidance for this work. 

\section*{Data Availability}

The observational data utilised in this article are publicly available at ISRO's Science Data Archive for \astrosat{} Mission (\url{https://astrobrowse.issdc.gov.in/astro_archive/archive/Home.jsp}). The additional information regarding data and data analysis will be available upon reasonable request.



\bibliographystyle{mnras}
\bibliography{xb1254} 




\appendix


\bsp	
\label{lastpage}
\end{document}